\documentclass[12pt]{article}
\usepackage{hyperref}
\usepackage{cite}
\usepackage{color}
\usepackage{graphicx}
\usepackage{amsmath}
\usepackage{amssymb}
\usepackage{xspace}

\def\baselinestretch{1.2}
\newcommand{\be}{\begin{equation}}
\newcommand{\ee}{\end{equation}}
\newcommand{\beq}{\begin{eqnarray}}
\newcommand{\eeq}{\end{eqnarray}}

\newcommand{\gone}[1]{{}}

\begin{document}
\begin{titlepage}
\begin{flushright}
MAD-TH-09-01
\end{flushright}

\vfil

\begin{center}

{\Large{\bf  Mass-spin relation for quark anti-quark bound states in non-commutative Yang-Mills theory}
}

\vfil

Sheikh Shajidul Haque and Akikazu Hashimoto

\vfil

Department of Physics, University of Wisconsin, Madison, WI 53706

\vfil

\end{center}

\begin{abstract}
\noindent
We investigate the relationship between mass and spin of quark
anti-quark bound state in non-commutative gauge theory with ${\cal
N}=4$ supersymmetry. In the large $N$ and large 't Hooft coupling
limit, these bound states correspond to rotating open strings ending
on a D-brane embedded into the supergravity background dual of the
non-commutative field theory. Although the physical configuration of
the open strings are drastically deformed by the non-commutativity, we
find that the relation between the energy and the angular momentum of
the bound state is unaffected by the non-commutativity parameter.  We
also clarify the holographic interpretation of quark anti-quark
potential in non-commutative gauge theories.

\end{abstract}
\vspace{0.5in}

\end{titlepage}
\renewcommand{\baselinestretch}{1.05}  

\section{Introduction}

The quark anti-quark potential, inferred from the expectation value of
Wilson loops, and the monopole anti-monopole potential, inferred from
the expectation value of the 't Hooft loops, are important physical
observables characterizing the phases of gauge field theories. Scaling
of Wilson or 't Hooft loop expectation value with respect to area
signals that a gauge theory is in an electrically, or magnetically,
confining phase, respectively. 


For the ordinary ${\cal N}=4$ supersymmetric Yang-Mills theory, there
is an elegant prescription for computing the quark anti-quark
potential using the dual type IIB $AdS_5 \times S_5$ description of
the theory \cite{Rey:1998ik,Maldacena:1998im}. The basic idea is to
introduce a D3-brane at some fixed radius $U_b$ in the Poincare
coordinate of the $AdS_5 \times S_5$ geometry and to interpret it as
corresponding to an $SU(N+1)$ theory broken to $SU(N) \times U(1)$.  A
fundamental string stretching from the D3 to the horizon of $AdS_5$
geometry corresponds to a W-boson charged as a fundamental with
respect to $SU(N)$ and also with respect to the $U(1)$. The mass of
such a state, which can be computed from the tension of the string
stretching from the D3 to the horizon, should match the vacuum
expectation value $U_b$ of the $U(1)$ component of the $SU(N+1)$
adjoint Higgs field.  Pushing the D3-brane to the boundary of $AdS_5$
corresponds to taking the infinite mass limit.  A configuration of an
infinitely massive quark and an anti-quark then corresponds to strings
with opposite orientations extending toward the boundary of the
$AdS_5$ geometry.  An important feature of this analysis is the fact
that as $U_b$ is sent to infinity keeping the configuration of the
strings in the small $U$ region constant, the distances between the
endpoints of the strings have a finite limit. This distance is the
distance separating the quark and the anti-quark. In this way, the
authors of \cite{Rey:1998ik,Maldacena:1998im} were able to compute the
potential between infinitely massive quark and an anti-quark as a
function of the distance separating the quark and the anti-quark, and
confirm the expected Coulomb behavior.

\begin{figure}
\centerline{\includegraphics{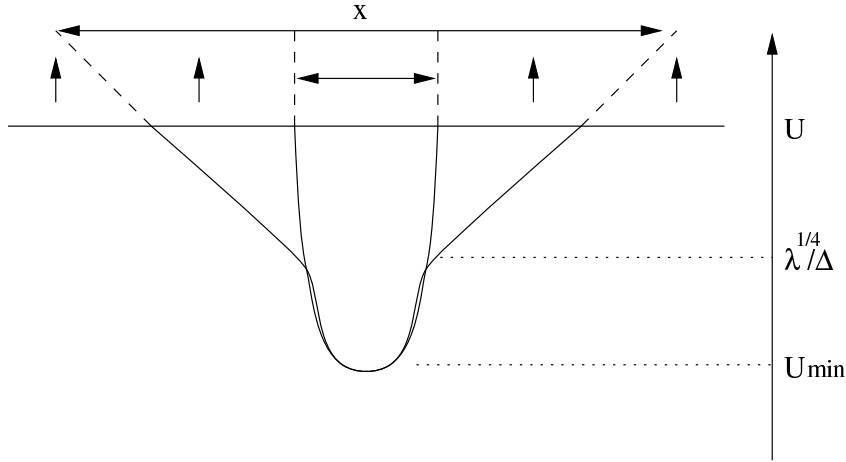}}
\caption{Configuration of string stretching from the boundary of
supergravity dual of non-commutative Yang-Mills theory. While for
$AdS_5 \times S_5$ the separation between two ends of the string
approaches to a finite value, it diverges in the case of the
non-commutative theory. This figure appeared originally in
\cite{si99}.  \label{figaa}}
\end{figure}

Soon after the construction of the supergravity dual for
non-commutative Yang-Mills theory, it was realized that this method of
computing quark anti-quark potential fails when applied to the
supergravity dual of non-commutative ${\cal N}=4$ supersymmetric
Yang-Mills theory \cite{Hashimoto:1999ut,si99,Maldacena:1999mh}. The
problem stems from the fact that in the large $U_b$ limit, the
distance separating the endpoint of the quarks did not approach a
finite limit \cite{si99,Maldacena:1999mh}, as we illustrate in figure
\ref{figaa} and review in section \ref{ozsec}.

Despite several attempts, e.g.\
\cite{si99,Maldacena:1999mh,Alishahiha:1999ci}, to make sense of this
strange behavior of quark anti-quark potential in non-commutative
field theories, its true physical meaning remains a
mystery.\footnote{Reference \cite{Alishahiha:1999ci} made an
interesting observation that if the quark anti-quark pair is moving in
the direction transverse to the direction separating the quarks at a
specific velocity $v$ (which depends on the distance separating the
quark and anti-quark) the potential can be computed along conventional
lines. However, the physical significance of the velocity $v$ invovled
in the computation was not obvious. We will provide the physical
significance of this velocity $v$ in section \ref{ozsec}.}  This is
rather unfortunate in light of the fact that quark anti-quark
potential is an observable which, by adjusting the distance between
the quark and the anti-quark, can probe the structure of space at
various length scales.\footnote{The holographic prescription for
computing the correlation function of gauge invariant operators was
also found at first to be rather subtle, but its status was clarified
in \cite{Das:2000md,Gross:2000ba}.}

In this article, we consider an alternative approach for probing the
quark anti-quark force. Specifically, we consider the spectrum of
spinning W-boson/anti-W-boson pair keeping $U_b$, the mass of the
W-boson, fixed in the process. The mass of the bound state as a
function of spin is an indirect measure of the attractive potential
via the virial theorem. In the dual supergravity formalism, these
states correspond to rotating open strings ending on a D3-brane placed
at $U=U_b$. When the spin of the bound state is taken to be large, the
open string can be treated semi-classically. The physical setup is
therefore quite similar to that of rotating folded closed strings
considered in \cite{Gubser:2002tv}.  Just as it was the case for
\cite{Gubser:2002tv}, one can rely on the semi-classical picture to
accurately capture the dynamics of the theory in the large $N$ and
large $\lambda$ limit.

This note is organized as follows. In section \ref{sec2}, we will
review the configuration of a basic spinning W/anti-W bound-state for
$SU(N+1) \rightarrow SU(N) \times U(1)$ theory at large $N$ and large
't Hooft coupling in the dual $AdS_5$ picture, and find expected
relations between their masses, their spins, and their size. This will
provide the basis of comparison for the non-commutative case.  We then
repeat the same analysis for the non-commutative case in section
\ref{ncsec}.  Not surprisingly, we find that the physical
configuration of the rotating string is significantly deformed as a
result of turning on the non-commutativity parameter. Despite this
drastic effect, however, we find that the relation between mass and
the spin of the bound state, as well as the distance between the
endpoints of the open strings, are unaffected by the non-commutative
deformation.  This is an unexpected result, possibly suggestive of
some hidden integrable structure which we have yet to identify. It
also suggests that the quark anti-quark force is unaffected by the
non-commutativity parameter in spite of the puzzle presented in
\cite{si99,Maldacena:1999mh}. 

We describe, in section \ref{ozsec}, a simple interpretation of the
moving $q \bar q$ pair of \cite{Alishahiha:1999ci} as the
configuration of vanishing linear momentum.

\section{Spinning open strings in $AdS_5 \times S_5$\label{sec2}}

In this section, we describe a configuration of rotating open strings
describing a bound state of W/anti-W pair.  Let us begin by reviewing
the rotating folded string solution in Minkowski space. Consider a Minkowski space in 2+1 dimensions parametrized in polar coordinates so that the metric has the form
\be ds^2 = -dt^2 + dr^2 + r^2 d \phi^2\ .  \ee
To describe a rigid string rotating around the origin at angular
velocity $\omega$, we can change coordinates to the co-rotating frame
where
\be ds^2 = -dt^2 + dr^2 + r^2  (d\phi + \omega dt)^2\ .  \ee
The rotating string, in this frame, is extended along the $t$ and the
$r$ coordinates, and localized at fixed $\phi$.  The world sheet
action is then
\be L =  T \int d \tau d \sigma\sqrt{(1 - \omega^2 r(\sigma)^2){dt(\tau) \over d \tau} {d r (\sigma) \over d \sigma}}  , \qquad T = {1 \over 2 \pi \alpha'} \ . \ee
Fixing the world sheet reparameterization gauge so that
\be t(\tau) = \tau, \qquad r(\sigma) = \sigma \ee
the action becomes
\be L =  -T \int d t d r \sqrt{(1 - \omega^2 r^2)}\ .  \ee
The $r$ has the range $-1/\omega < r < 1/\omega$. The pieces of the
string at the $r=\pm 1/\omega$ are moving at the speed of light. It is
at this point that the string folds back on itself. One can compute
the energy and the spin of this configuration
\be E = 2 \int_{-1/\omega}^{1/\omega} dr \, \left({\partial L \over \partial \omega} \omega - L \right) = 2\pi {T \over  \omega} \label{energy} \ee
\be S =
2 \int_{-1/\omega}^{1/\omega} dr {\partial L \over \partial \omega} =   \pi {T \over \omega^2} \label{spin} \ee
from which we conclude
\be E = 2 \sqrt{\pi T S} \ . \label{esrigid} \ee
These are the semi-classical description of strings in the leading
Regge trajectory in Minkowski space-time.

Embedding these semi-classical folded closed string solutions into
$AdS_5 \times S_5$ geometry is relatively straight forward. They were
studied first in \cite{Gubser:2002tv} and were interpreted as
corresponding to an operator with large spin and small twist. The
semi-classical analysis was shown to be reliable in the limit of large
$\lambda$ and $N$.

Now, let us consider similar construction for open strings ending on a
D3-brane in $AdS_5 \times S_5$.  This discussion is essentially a
review of the of the analysis of open strings ending on a D7-brane,
considered in \cite{Kruczenski:2003be}, whose dynamics is
identical. Consider the metric of $AdS_5 \times S_5$
\be ds^2 = \alpha' \left( {U^2 \over \sqrt{\lambda}} (-dt^2 + dx_1^2 + dr^2 + r^2 d \phi^2) + {\sqrt{\lambda} \over U^2} dU^2 + \sqrt{\lambda} d \Omega_5^2\right) \label{metric}\ee
with a D3-brane located at some fixed point in $S_5$ and at $U=U_b$
and extend along the $t$, $r$, $\phi$, and $x_1$ directions. This is a
familiar configuration corresponding to the $SU(N+1)$ gauge theory
broken to $SU(N) \times U(1)$. For rigid rotating strings in the
co-rotating frame, the world sheet action takes the form
\be L = {1 \over 2 \pi} \int d\tau\, d\sigma \,  {d t(\tau) \over d \tau} {U(\sigma)^2 \over \sqrt{\lambda}} \sqrt{(1 - \omega^2 r(\sigma)^2)\left(\left({d r(\sigma) \over d \sigma}\right)^2 + {\lambda \over U(\sigma)^4} \left({\partial U(\sigma) \over \partial \sigma}\right)^2\right)}\ .  \label{action} \ee
Note that the dependence on $\alpha'$ has canceled out in
(\ref{action}) as is expected for a dual of a gauge field theory.

There are several manipulations one can do to transform this action
into a more manageable form.  Let us introduce a scale $U_0$. This
simply provides a reference for quantifying other dimensionful
quantities, but does not, by itself, break the conformal invariance of
the underlying physics.

We can then define dimensionless parameters
\be U(\sigma) = \lambda^{1/2} {U_0 \over x(\sigma)}, \qquad r(\sigma) = U_0^{-1} y(\sigma), \qquad \omega = U_0 w \ee
and work in temporal gauge
\be t = U_0^{-1} \tau \ee
to bring the action into the form
\be L = -{\lambda^{1/2} \over 2 \pi} \int d \tau \, d \sigma \, {1 \over x(\sigma)^2} \sqrt{(1 - w^2 y(\sigma)^2) (x'(\sigma)^2+ y'(\sigma)^2)} \ . \label{action2} \ee

We have not yet fixed the gauge with respect to reparameterization of
$\sigma$. One way to do this is to introduce a Lagrange multiplier
\be L = -{\lambda^{1/2} \over 2 \pi} \int d \tau \, d \sigma \, {1 \over x(\sigma)^2} \sqrt{(1 - w^2 y(\sigma)^2) } \left({e \over 2} + {1 \over 2 e} (x'(\sigma)^2+ y'(\sigma)^2)\right)\ee
and impose the gauge condition
\be e = {1 \over x(\sigma)^2} \sqrt{(1 - w^2 y(\sigma)^2)}\ . \label{gaugecond}\ee
Then, the action takes the form
\be L = - {\lambda^{1/2} \over 2 \pi} \int d \tau \, d \sigma \, \left({1 \over2 } x'(\sigma)^2 + {1 \over 2} y'(\sigma)^2  - V(x(\sigma),y(\sigma)) \right),\ee
where
\be  V(x(\sigma),y(\sigma)) = -{1 - w^2 y(\sigma)^2 \over 2 x(\sigma)^4}\ .  \label{pot}\ee
Note, at the level of classical equations of motion, that this action
can be interpreted as a trajectory of a particle in 2 dimensions with
its position parametrized by $x(\sigma)$ and $y(\sigma)$, evolving in
time variable parametrized by $\sigma$, under the influence of a
static potential $V(x,y)$.

The equation of motion with respect to variation of $e$ implies a constraint
\be \sqrt{x'(\sigma)^2 + y'(\sigma)^2} = e.  \label{constraint}\ee
Combining  (\ref{gaugecond}) and (\ref{constraint}) implies
\be 
H = {1 \over 2} x'(\sigma)^2 + {1 \over 2} y'(\sigma)^2 +V(x,y) = 0 \ee
which can be interpreted as an additional constraint that the total
energy is set to zero.

To proceed with the analysis of open strings ending on a D3-brane,
note that a surface defined by $U = U_b$ is mapped to
\be x = {U_0 \over U_b} \equiv x_b \ .  \ee
In light of the conformal invariance of the ${\cal N}=4$ theory, one
should think of either $U_b$ or $\omega$ as introducing a scale to the
analysis, and their dimensionless ratio
\be {\omega  \over U_b} = \lambda^{1/2} w x_b \ee
as being physically meaningful.

We are interested in finding a steady-state configuration of open
strings with endpoints on $x=x_b$. The presence of a D3-brane imposes
a Dirichlet boundary condition for the $x(\sigma)$ variable and a
Neumann boundary condition for the $y(\sigma)$. In other words, we have
\be x(\sigma=0) = x_b, \qquad y'(\sigma=0) = 0 \ee
at one end of the string. For $\sigma = \sigma_{max}$, we also require
\be x(\sigma=\sigma_{max}) = x_b, \qquad y'(\sigma=\sigma_{max}) = 0 \ . \ee

The initial condition $y(\sigma=0)$ could a priori take on any
value. However, the constraint $H=0$ imposes a condition that
\be x'(\sigma=0) = \pm {\sqrt{1 - w^2 y(0)^2} \over x(0)^2} \ . \ee
The data $x(0)$, $x'(0)$, $y(0)$, and $y'(0)$ is sufficient to
determine the solution to the equation of motion for $x(\sigma)$ and
$y(\sigma)$, but one does not expect in general for the boundary condition
to be satisfied at $\sigma=\sigma_{max}$. Once $x_b$ is fixed, the
only adjustable parameter is $y(0)$. One can consider varying $y(0)$
and search for special values at which the boundary condition at
$\sigma=\sigma_{max}$ is also satisfied. One can visualize this
process as that of shooting a pin-ball at $x=x_b$ and $y=y(0)$ in a
potential $V(x,y)$ and requiring that the ball comes back to $x=x_b$
at $\sigma = \sigma_{max}$ at an appropriate angle defined by the
boundary conditions.  Since we are imposing one condition on one
variable, the expectation is that there is at most a discrete set of
solutions.

The equations of motion for $x(\sigma)$ and $y(\sigma)$ are second
order, coupled, and non-linear. For $w=0$, one can solve the equations
analytically and reproduce the analysis of
\cite{Rey:1998ik,Maldacena:1998im}, but for generic values, we were
forced to solve the equations numerically. The rest of this section describes  
the general result of this numerical analysis.

Without loss of generality, we can set $x_b=1$.  While different
values of $w$ are physically distinct, let us set $w=1$ for the sake
of concreteness. One can vary $y(0)$ in the range $-1/w < y(0) < 1/w$
and satisfy the constraint $H=0$. For each possible value of $y(0)$,
we solve the equation of motion, and look for $\sigma_{max}>0$ for
which $x_b=1$. One can then compute $y'(\sigma_{max})$ and plot this
against $y(0)$. The zeros of $y'(\sigma_{max})$ as a function of
$y(0)$ corresponds to a solution of the equation of motion satisfying
all of the boundary conditions.

\begin{figure}[t]
\centerline{\includegraphics[width=3.5in]{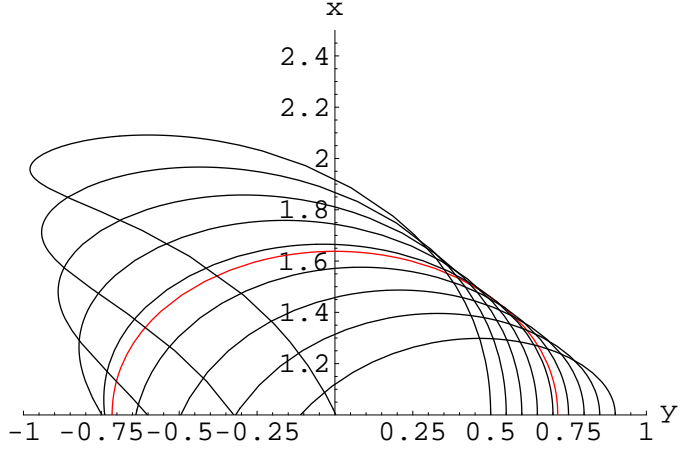}}
\caption{Parametric plot of the solutions $(x(\sigma),y(\sigma))$ with
initial conditions $x(0)=1$, $y'(0)=0$, and fixed $y(0)$ in the range
$0.5 < y(0) < 0.9.$ \label{figa}}
\end{figure}

For the sake of illustration, set of solutions $(x(\sigma),y(\sigma))$
with initial conditions $x(0)=1$, $y'(0)=0$, and fixed $y(0)$ in the
range $0.5 < y(0) < 0.9$ are displayed in figure \ref{figa}. Note that
$y'(\sigma_{max})$ varies as $y(0)$ is varied. There is a special
value of $y(0)\approx 0.715$, illustrated in red curve, for which
$y'(0)=0$ satisfying the boundary condition at $\sigma =
\sigma_{max}$.  This is a consistent spinning open string
configuration in $AdS_5 \times S_5$.

\begin{figure}[t]
\centerline{\includegraphics[width=3.5in]{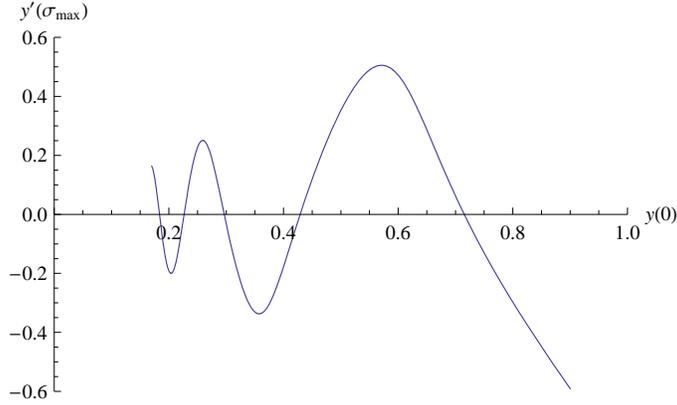}}
\caption{$y'(\sigma_{max})$ for the solution of equation of motion as a function of the initial condition $y(0).$\label{figb}}
\end{figure}

To be more systematic, one can plot $y'(\sigma_{max})$ for the
solutions of the equation of motion as a function of the initial condition
$y(0)$. This is illustrated for the range of $0.175<y(0)<0.9$ in
figure \ref{figb}. We stopped at $y(0) \approx 0.2$ since numerical
load becomes heavier for smaller value of $y(0)$. The zero of this
function near $y(0)\approx 0.715$ is readily apparent. Also apparent
are other zeros at $y(0)$ taking approximate values 0.184, 0.227,
0.297, 0.429.  The significance of these other zeros and the behavior
for smaller values of $y(0)$ will be discussed in the appendix \ref{appa}.

\begin{figure}
\centerline{\includegraphics[width=3.5in]{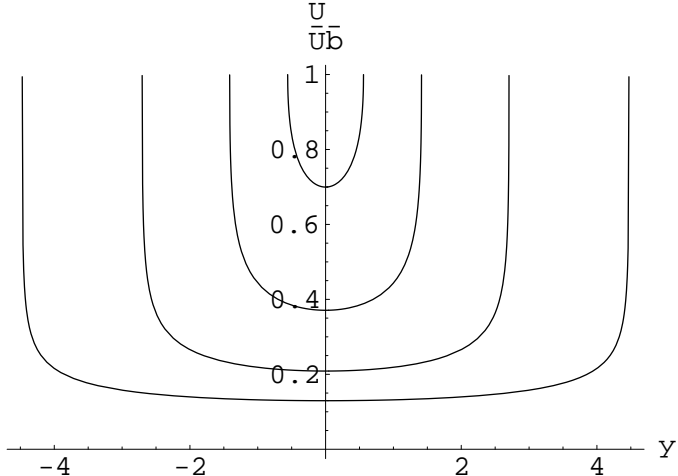}}
\caption{Solution of equation of motion and the boundary condition for D3-branes placed at $U=U_b$, illustrated as $U/U_b$ as a function of $y$ for various values of $w$. The configuration with large $y$ corresponds to small $w$ and large $S$. \label{figc}}
\end{figure}

It would be instructive to explore how these solutions change as the
values of $w$ is varied keeping $x_b$ fixed. This is illustrated in
figure \ref{figc} where we have chosen to plot $U/U_b$ instead of $x$
as a function of $y$ to facilitate the comparison with configurations
described in \cite{Rey:1998ik,Maldacena:1998im}.  The curves with
larger values of $y$ corresponds to smaller values of $w$. From figure
\ref{figc}, it is apparent that most of the energy comes from the rest
masses of W and anti-W when they are separated by  large distances with
small values of $w$.

\begin{figure}
\centerline{\includegraphics[width=3.5in]{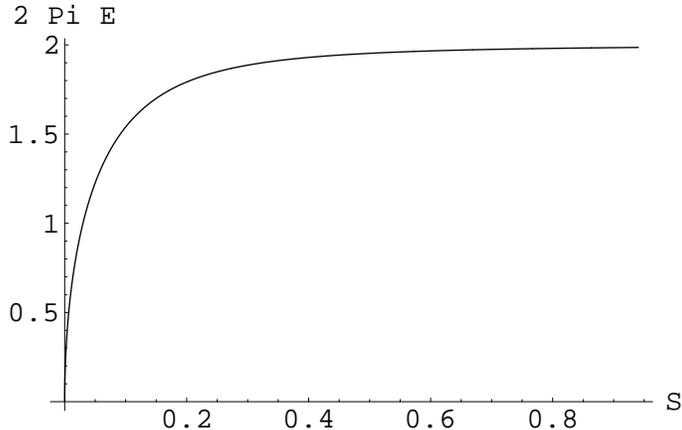}}
\caption{Plot of energy $E$ v.s.\ angular momentum $S$ for a positronium-like bound state of W/anti-W pair in a strongly coupled ${\cal N}=4$ theory with gauge group $SU(N) \times U(1)$. \label{figd}}
\end{figure}

A very useful exercise is to plot $E$ as a function of $S$ by
substituting the solution to (\ref{energy}) and (\ref{spin}). The
result of this analysis is illustrated in figure \ref{figd}. For small
values of $S$, we find
\be E \sim \sqrt{\frac{\pi T S}{2}} \ee
which has the form of the relation (\ref{esrigid}) for rotating rigid
string in flat space, with the effective tension $T= {U_b^2 / 2\pi
\lambda^{1/2}}$ for  strings extended along $U=U_b$.  For
large $S$, the energy asymptotes to twice the rest mass of the
W-bosons. There is a factor of $\sqrt{2}$ difference between
(\ref{esrigid}) since the open strings are not folded. (This gives
rise to factors of 1/2 in $E$ and $S$.)

These semi-classical steady state rotating open string solutions in
$AdS_5 \times S_5$ provide the main tool for the subsequent analysis
in the remainder of this paper. They are positronium-like bound state
of W and an anti-W. In stating such a result, one should keep in mind
that the semi-classical analysis of the type presented here is subject
to world sheet ($\alpha'$) and space-time ($g_s$) quantum
corrections. Indeed, rigid rotating strings typically decay via
fragmentation and emission of closed strings
\cite{Mitchell:1988qe,Wilkinson:1989tb,Mitchell:1989uc,Okada:1989sd,Balasubramanian:1996xf}. Fortunately,
as was the case for \cite{Gubser:2002tv}, the strong 't Hooft coupling
and large $N$ limit allows one to treat the semi-classical analysis to
be reliable quantitatively, in that the quantum corrections are small.

The fact that there are positronium-like bound states in the spectrum
of $SU(N+1) \rightarrow SU(N) \times U(1)$ theory implies that they
should appear as resonances, for example, in $2 \rightarrow 2$
scattering process. Recently, there have been significant efforts
directed toward computations of scattering amplitudes taking advantage
of the unique features of ${\cal N}=4$ supersymmetry. See
e.g. \cite{Dixon:1996wi,Cachazo:2005ga,Dixon:2008tu} for reviews and
references to recent developments. Perhaps some of these techniques
can be used to confirm the rich spectrum of ${\cal N}=4$ theory in the
Coulomb branch.

The rigid rotating strings in the treatment of \cite{Gubser:2002tv}
were steady state solution with respect to global time of $AdS_5
\times S_5$. This allowed their mass to be compared to the scaling
dimension of gauge invariant operators. The fact that the positronium-
like state found in this note is a steady state with respect to
Poincare time makes similar interpretation somewhat subtle. One way to
overcome this obstacle might be to consider open strings ending on
giant gravitons blown up in the $AdS_5$
\cite{Grisaru:2000zn,Hashimoto:2000zp}, along the lines of
\cite{Berenstein:2005vf,Berenstein:2005fa,Berenstein:2006qk}.  Open
spinning strings in $AdS_5 \times S_5$ and related models were also
considered in
\cite{Bigazzi:2004ze,Kruczenski:2004me,Okamura:2005cj,Mann:2006rh}.

In the following section, we examine how the analysis of the present section is modified when  the non-commutativity parameter is non-vanishing.

\section{Positronium in non-commutative SYM\label{ncsec}}

The supergravity background corresponding to non-commutative Yang-Mills theory has a relatively simple form  \cite{Hashimoto:1999ut,si99,Maldacena:1999mh} whose metric and the NSNS 2-form are given by
\beq ds^2 &=& \alpha' \left( {U^2 \over \sqrt{\lambda}} \left(-dt^2 + dx_1^2+ {1 \over 1 + \lambda^{-1} \Delta^4 U^4} (dr^2 + r^2 d \phi^2) \right) + {\sqrt{\lambda}\over U^2} (dU^2+U^2 d \Omega_5^2\right) \label{sugra1}\cr
B &=& \alpha'{\Delta^2 U^4 \over \lambda + \Delta^4 U^4} r dr \wedge d \phi\ .  \eeq

As we reviewed in the introduction, the status of Wilson loop
observables along the lines of \cite{Rey:1998ik,Maldacena:1998im}
remains unclear, in light of the fact that the world sheet
configuration $r(U)$ with finite $U_{min}$ does not take finite value
when $U \rightarrow \infty$ \cite{si99,Maldacena:1999mh}.  Regarding
this issue, it is worth noting an interesting observation made in
\cite{Alishahiha:1999ci} that $r(U=\infty)$ can be made to take on
finite value if the strings were made to move at a special fixed
velocity, depending on the distance separating the endpoints of the
strings at $U = \infty$, along the non-commutative plane. The physical
significance of the velocity required for the $q \bar q$-pair is not
immediately obvious.\footnote{We will suggest an interpretation for
the physical significance of the specific velocity involved in the
construction of \cite{Alishahiha:1999ci} in section \ref{ozsec}.} The
study of positronium-like bound states of W and anti-W would,
therefore, provides an interesting new perspective on the nature of
interactions between opposite charges in non-commutative field
theories.  Our analysis of the positronium-like bound state will
suggest a natural interpretation for the velocity of the moving
strings used in \cite{Alishahiha:1999ci} which we will elaborate in
section \ref{ozsec}.

Let us begin by considering a steady state rotating ansatz for the
strings.  The action for the world sheet will then take the form
\beq L &=& \int d \tau \, d \sigma \, {U(\sigma)^2 \over \sqrt{\lambda}} \sqrt{
\left( 1 - {\omega^2 r(\sigma)^2 \over 1 + \lambda^{-1} \Delta^4 U(\sigma)^4}\right)
 \left({r'(\sigma)^2 \over 1 + \lambda^{-1} \Delta^4 U(\sigma)^4} + {\lambda U'(\sigma)^2 \over U(\sigma)^4}\right)} \cr
&& \qquad + {\omega \Delta^2 r(\sigma) r'(\sigma) U(\sigma)^4 \over \lambda + \Delta^4 U(\sigma)^4}\ .  
\eeq
Just as we did before, it is useful to rescale 
\be U(\sigma) = {\lambda^{1/2} U_0 \over x(\sigma)} , \qquad r(\sigma) = {1 \over U_0}y(\sigma), \qquad \omega = U_0 w  , \qquad \Delta = {\lambda^{-1 / 4} \over U_0} q\ee
and introduce a Lagrange multiplier $e$ so that the action becomes
\be L = {f\over 2} (e + {1 \over e} \left(\left(1 + {q^4 \over x(\sigma)^4}\right) x'(\sigma)^2 + y'(\sigma)^2\right) + {q^2 w y(\sigma) y'(\sigma) \over q^4 + x(\sigma)^4}\ee
where
\be f^2 = {q^4 + x(\sigma)^4 (1 - w^2 y(\sigma)^2) \over (q^4 + x(\sigma)^4)^2} \ . \ee
With the gauge condition
\be e = f \ee
the action will take the form
\be L =  {1 \over 2} \left(\left(1 + {q^4 \over x(\sigma)^4}\right) x'(\sigma)^2 + y'(\sigma)^2\right) - V\left(x(\sigma),y(\sigma)\rule{0ex}{2ex}\right)+ A_y\left(x(\sigma),y(\sigma)\rule{0ex}{2ex}\right) y'(\sigma)\ee
\be V\left(x(\sigma),y(\sigma)\rule{0ex}{2ex}\right) = -{f^2 \over 2} ,\qquad 
A_y\left(x(\sigma),y(\sigma)\rule{0ex}{2ex}\right) = {q^2 w y(\sigma)  \over q^4+ x(\sigma)^4} \ , \ee
with a constraint
\be H = {1 \over 2} \left(\left(1 + {q^4 \over x(\sigma)^4}\right)x'(\sigma)^2 + y'(\sigma)^2\right)- {1 \over 2} f^2 = 0 \ . \ee

In order to make this action look more like a motion of a particle in a potential, one can further change variables 
\be z(x) = 1  + \int_1^x  dx_0 \left(1+{q^4 \over x_0(\sigma)^4}\right)^{1/2} \ee
so that the action is
\be L =  {1 \over 2} \left( z'(\sigma)^2 + y'(\sigma)^2\right) - V\left(z(\sigma),y(\sigma)\rule{0ex}{2ex}\right)+ A_y\left(z(\sigma),y(\sigma)\rule{0ex}{2ex}\right) y'(\sigma)\ . \ee
Note that $z=0$ does not have any special significance. The boundary
points $x=0$ are mapped to $z=-\infty$. This action describes a point
particle moving in a $(z,y)$ plane with potential $V(z,y)$ in the
presence of a magnetic field given by a vector potential
$A_y(z,y)$. Unlike in the case of the constant magnetic field, the
$A_y$ term, which came from the NSNS $B$-field, does affect the
equation of motion. It also changes the relation between $z'(\sigma)$
and its conjugate momentum. However, it does not affect the
conservation of the Hamiltonian
\be H = {1 \over 2} \left( z'(\sigma)^2 + y'(\sigma)^2\right) + V\left(z(\sigma),y(\sigma)\rule{0ex}{2ex}\right) \ . \ee

\begin{figure}
\centerline{\includegraphics[width=4in]{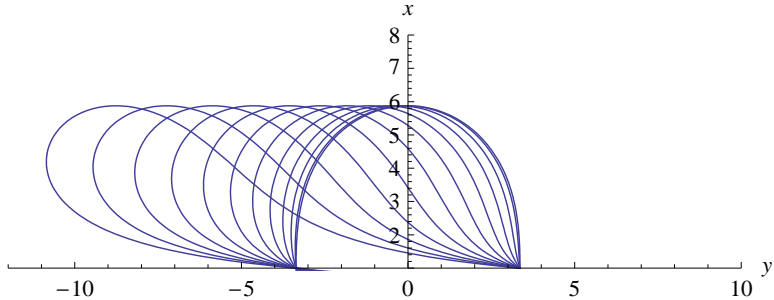}}
\caption{Positronium-like bound state of W/anti-W pair in dual supergravity description of non-commutative supersymmetric Yang-Mills theory. Here, we have taken $0 \le q \le 5.5$ and $\omega=0.05.$\label{figh}}
\end{figure}

For the purpose of solving the equations of motion numerically, the $(x,y)$ coordinates suffice. The boundary condition for the $y(\sigma)$ for this case is
\be y'(\sigma) +{q^2 w y(\sigma) \over q^4 + x(\sigma)^4} = 0 \ . \ee
For the purpose of illustration, let us consider setting $w=0.05$,
$x_b=1$, and $q$ ranging from $q=0$ to $q=5.5$. The solutions
corresponding to this case is illustrated in figure \ref{figh}. 

One notable feature of the solutions illustrated in figure \ref{figh}
is the fact that the position of the string end points $y(0)$ and
$y(\sigma_{max})$, as well as the maximum $x_{max}$, appears to be
unaffected by changes in $q$. This may well be an artifact of probing
only narrow range of parameters. Nonetheless, the independence of these data on $q$ seems remarkably robust and is suggestive that there might be some hidden symmetry, although we were unable to identify one.

Note that as the non-commutativity is increased, the boundary
condition will tilt the string as it ends on the brane, causing a
narrow fluxtube to appear before the strings penetrate the bulk AdS
region, indicating the spreading of the flux.  This is compatible with
the intuition that the positions of the quarks are becoming more
uncertain as $q$ is increased
\cite{si99,Maldacena:1999mh}. Nonetheless, the existence of these
solutions indicates that a W/anti-W bound state do exist. One can
further study the relation between energy and spin along the lines of
what we illustrated in figure \ref{figd}.  We have repeated this
analysis of determining $E$ as a function of $S$ for various fixed
values of $q=0.5$, $q=1$ and $q=5$. Rather remarkably, we found that
the relation $E(S)$ is unaffected by the changes in $q$.\footnote{This
result is qualitatively different from the result reported in figures
6 and 7 of \cite{Arean:2005ar}. The source of this discrepancy is
explained in the note added.}  This may also be an indication that
there is some hidden symmetry which we have not yet accounted.

\begin{figure}
\centerline{\includegraphics[width=3.5in]{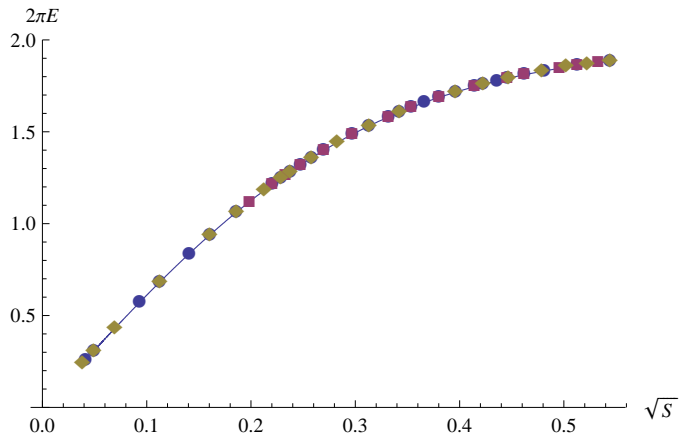}}
\caption{Numerical computation of $E(S)$ for $q=0.5$, $q=1.0$ and $q=5.0$ superposed on the $q=0$ curve. \label{figi}}
\end{figure}

It should be noted that this analysis closely parallels that of
\cite{Arean:2005ar} which also considered the dynamics of rotating
open strings ending on a D7-brane. In fact the equation of motion and
the solution we present are identical. Yet, we reach a drastically
different conclusion, namely that $E(S)$, illustrated in figure
\ref{figi}, is {\it insensitive} to the effects of non-commutative
deformation. The cause of this apparent discrepancy is the fact that
the angular momentum computed in \cite{Arean:2005ar} did not include
the contribution from the Kalb-Ramond term of the world sheet
action. Kalb-Ramond terms certainly contribute to the canonical
momentum and angular momentum for the dynamics of open strings
\cite{Bigatti:1999iz}; and with the contribution from the Kalb-Ramond
included, we find compelling numerical indication that $E(S)$ is
independent of the non-commutativity deformation.  This is an
important conclusion since it implies, in spite of the puzzling
behavior of the Wilson lines \cite{si99,Maldacena:1999mh}, that the
quark anti-quark potential is unaffected by the non-commutative
parameter.

\section{Canonical momentum of the moving $q \bar q$ pair \label{ozsec}}

In the analysis of the positronium-like bound state, we found that the
relation between the energy $E$ and the canonical angular momentum
$S$ which includes the contribution of the Kalb-Ramond term is
unaffected by the changes in the non-commutativity parameter.  This
observation offers a natural interpretation for the specific velocity
of the moving $q \bar q$ pair in the analysis of
\cite{Alishahiha:1999ci}.

Let us begin by recalling the construction of \cite{Alishahiha:1999ci}. A string world sheet embedded in the supergravity background (\ref{sugra1}) which we re-express in Cartesian coordinates \cite{Hashimoto:1999ut,si99,Maldacena:1999mh}
\beq
ds^2 & = & \alpha'\left\{ {U^2 \over \sqrt{\lambda}}(-dt^2 + dx_1^2) 
+ {\sqrt{\lambda} U^2 \over \lambda + U^4 \Delta^4} (dx_2^2+dx_3^2) 
+ {\sqrt{\lambda} \over U^2} dU^2 + \sqrt{\lambda} d \Omega_5^2 \right\} \cr
B_{23} & = &  {\alpha' \Delta^2 U^4 \over \lambda + \Delta^4 U^4}, 
\eeq
will minimize the action
\be S = \int dt \, dU \sqrt{-g_{00} \left(g_{UU}  + g_{33} \left({d x_3(U) \over dU}\right)^2\right)} \label{lag1} \ee
where we have taken the $q \bar q$ pair to be separated along the $x_3$ direction and stationary in the $x_1,x_2$ directions. If the $q \bar q$ pair moves in the $x_2$ direction with velocity $v$, the action (\ref{lag1}) is modified to take the form
\be S = \int dt \, dU \sqrt{-(g_{00}+g_{22} v^2) \left( g_{UU}  + g_{33} \left({d x_3(U) \over dU}\right)^2\right)}  - B_{23} v {d x_3(U) \over dU} \label{lag2}. \ee
As was the case in \cite{Rey:1998ik,Maldacena:1998im}, it is possible to constrain
\be {\delta S \over \delta x_3'(U)} = \mbox{fixed} \ee
which can be solved for $x_3'(U)$
\be x_3'(U) = c(v,U_{min}) + {\cal O}(U^{-4}) \ee
giving rise to the linearly growing $x_3(U)$ with the slope
$c(v,U_{min})$ depending on $v$ and $U_{min}$ illustrated in figure
\ref{figaa}.

The key observation in \cite{Alishahiha:1999ci} is the fact that by choosing $v$ to take a specific value
\be v = - {\Delta^2 U_{min}^2 \over \sqrt{\lambda}} \label{specialv} \ee
the slope constant $c(v,U_{min})$ vanishes and the asymptotic behavior of $x_3'(U)$ simplifies drastically to take the form
\be x_3'(U) = {\sqrt{\lambda} U_{min}^2  \over U^2 \sqrt{U^4 - U_{min}^4}} .\ee
This, in fact, is identical to the configuration of \cite{Rey:1998ik,Maldacena:1998im} at zero velocity and vanishing non-commutativity.

There is a natural interpretation of the special velocity
(\ref{specialv}) in light of the conclusion of the previous section.
If one considers the canonical momentum, one finds that it vanishes
identically, precisely when $v$ takes the value (\ref{specialv}).
\beq P_2 &=& \left. \int dU {d \over dv} \left( \sqrt{-(g_{00} +g_{22} v^2)\left( g_{UU}  + g_{33} \left({d x_3(U) \over dU}\right)^2\right)}  - B_{23} v {d x_3(U) \over dU}\right) \right|_{v = -\Delta^2 U_{min}^2/\sqrt{\lambda}}\cr
&=& \int dU\,  {\Delta^2 \sqrt{\lambda} U^2 U_{min}^2 \over \lambda + \Delta^4 U^4} \left( {1 \over \sqrt{U^4 - U_{min}^4}} - 
{1 \over \sqrt{U^4 - U_{min}^4}}\right) \cr
& = & 0 
\eeq
This clarifies the physical significance of the special value
(\ref{specialv}) for $v$. The quark anti-quark potential are well
defined in non-commutative plane if one fixes the momentum, instead of
the position, of the quarks in the direction orthogonal to the
direction separating the quarks. In order to fix the value of this
momentum to zero, one sets $v$ to the special value
(\ref{specialv}). For that value of $v$, the quark anti-quark
potential has a smooth commutative limit. In fact, the potential is
insensitive to the non-commutativity parameter. This observation
further emphasizes the importance of considering the canonical
momentum including the contribution of the Kalb-Ramond term.

\section{Conclusions}

In this article, we considered the dynamics of W-boson and
anti-W-boson pair in ${\cal N}=4$ supersymmetric non-commutaitive
Yang-Mills theory in 3+1 dimensions, with spontaneously broken gauge
group $SU(N+1)\rightarrow SU (N)\times U(1)$. We studied this system
at large $N$ and large 't Hooft coupling, where the dual supergravity
description in terms of a probe D3-brane embedded into an $AdS_5
\times S_5$ background in type IIB string theory is valid. In this
description, the W-bosons are represented as strings stretching from
the D3-brane to the horizon of $AdS_5$.



The question of how to correctly formulate quark anti-quark potential
as a holographic observable, along the lines of
\cite{Rey:1998ik,Maldacena:1998im} for the supergravity dual of
non-commutative field theory, has been a long standing puzzle
\cite{si99,Maldacena:1999mh}. It is, therefore, quite interesting that
positronium-like bound states of W/anti-W pair can be shown to exist
in non-commutative field theories at large $N$ and large 't Hooft
coupling, and for their energy and spin to be completely insensitive
to the non-commutativity parameter, as we illustrated in figure
\ref{figi}.  In order to arrive at this conclusion, it is important to
consider the canonical angular momentum including the contribution of
the Kalb-Ramond term. Viewing $q\bar q$ potential as being at fixed
momentum, as opposed to fixed position in the direction transverse to
the direction separating the quarks, also gives rise to a simple
result independent of the non-commutativity parameter. By using the
mixed position space/momentum space quantum numbers as a label, the
holographic interpretation of the Wilson loop observables in
non-commutative gauge theories appears to clarify significantly.

\section*{Acknowledgements}

We would like to thank Koji Hashimoto, Horatiu Nastase, and Peter Ouyang for discussions and Angel Paredes for bringing \cite{Kruczenski:2003be,Arean:2005ar} to our attention and further correspondences.
This work was supported in part by the DOE grant DE-FG02-95ER40896.

\appendix

\section{Excited positronium-like bound states\label{appa}}

\begin{figure}[t]
\begin{tabular}{cc}
\includegraphics[width=3in]{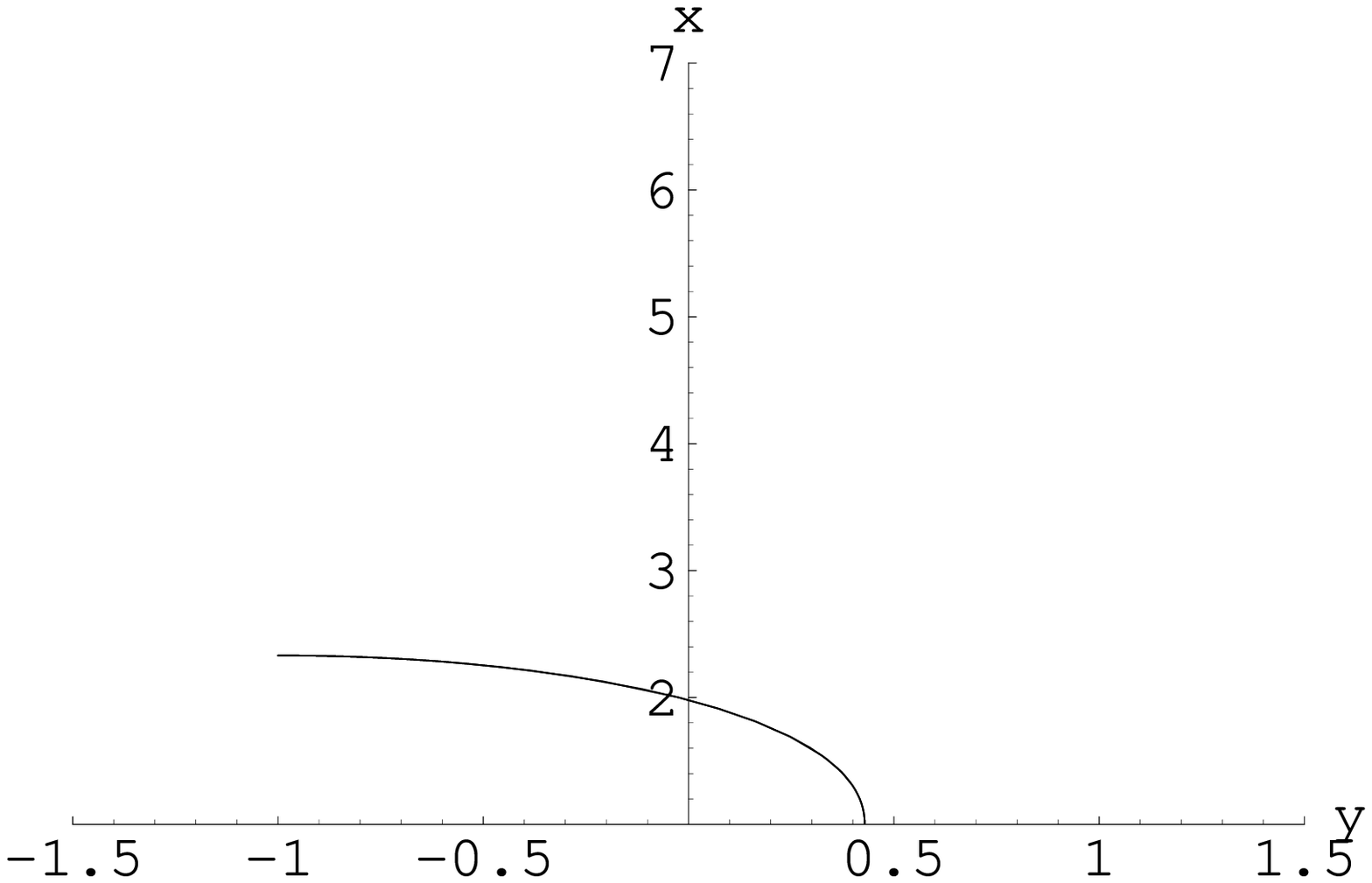} & 
\includegraphics[width=3in]{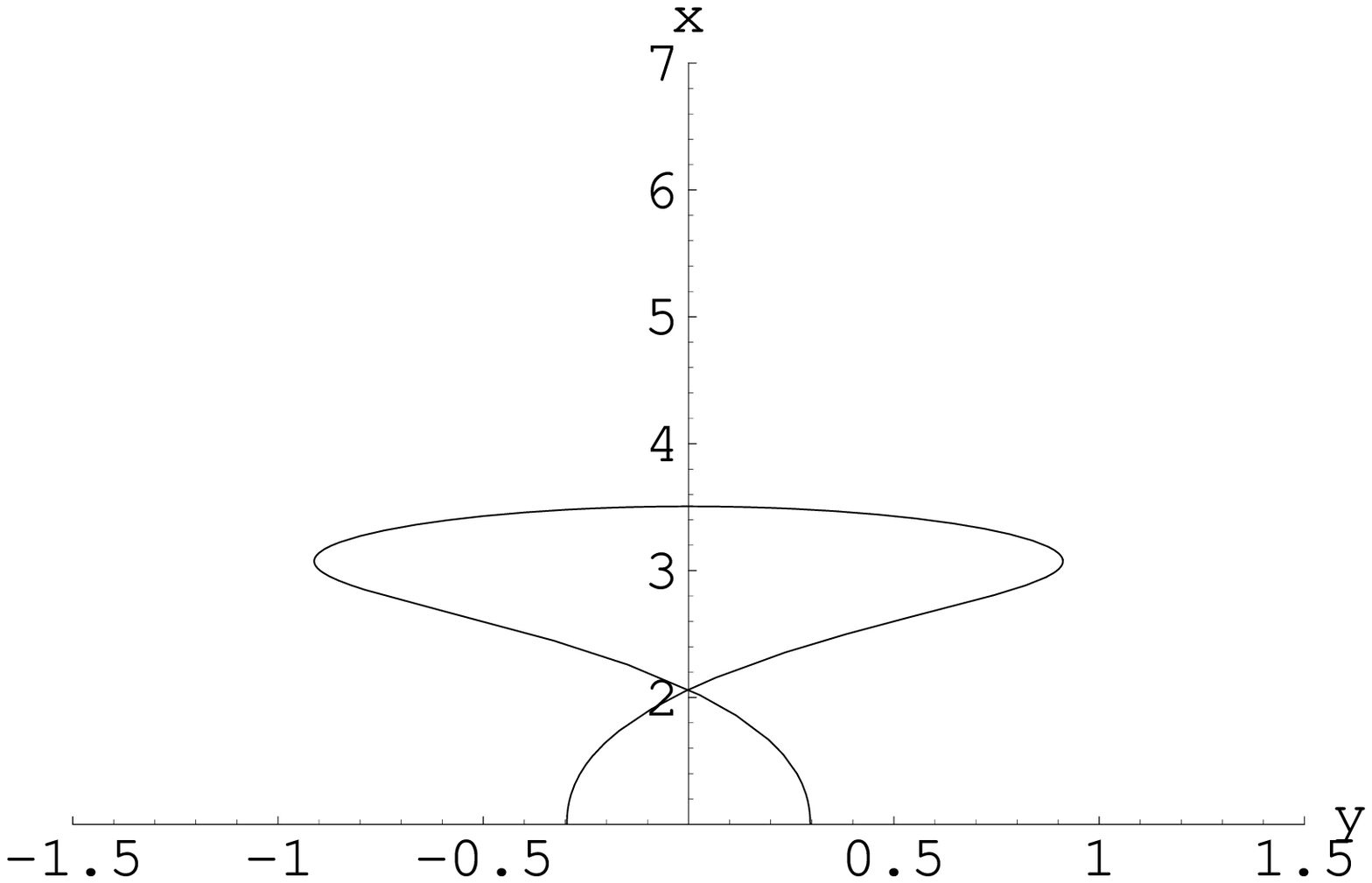}\\
(a) & (b) \\
\includegraphics[width=3in]{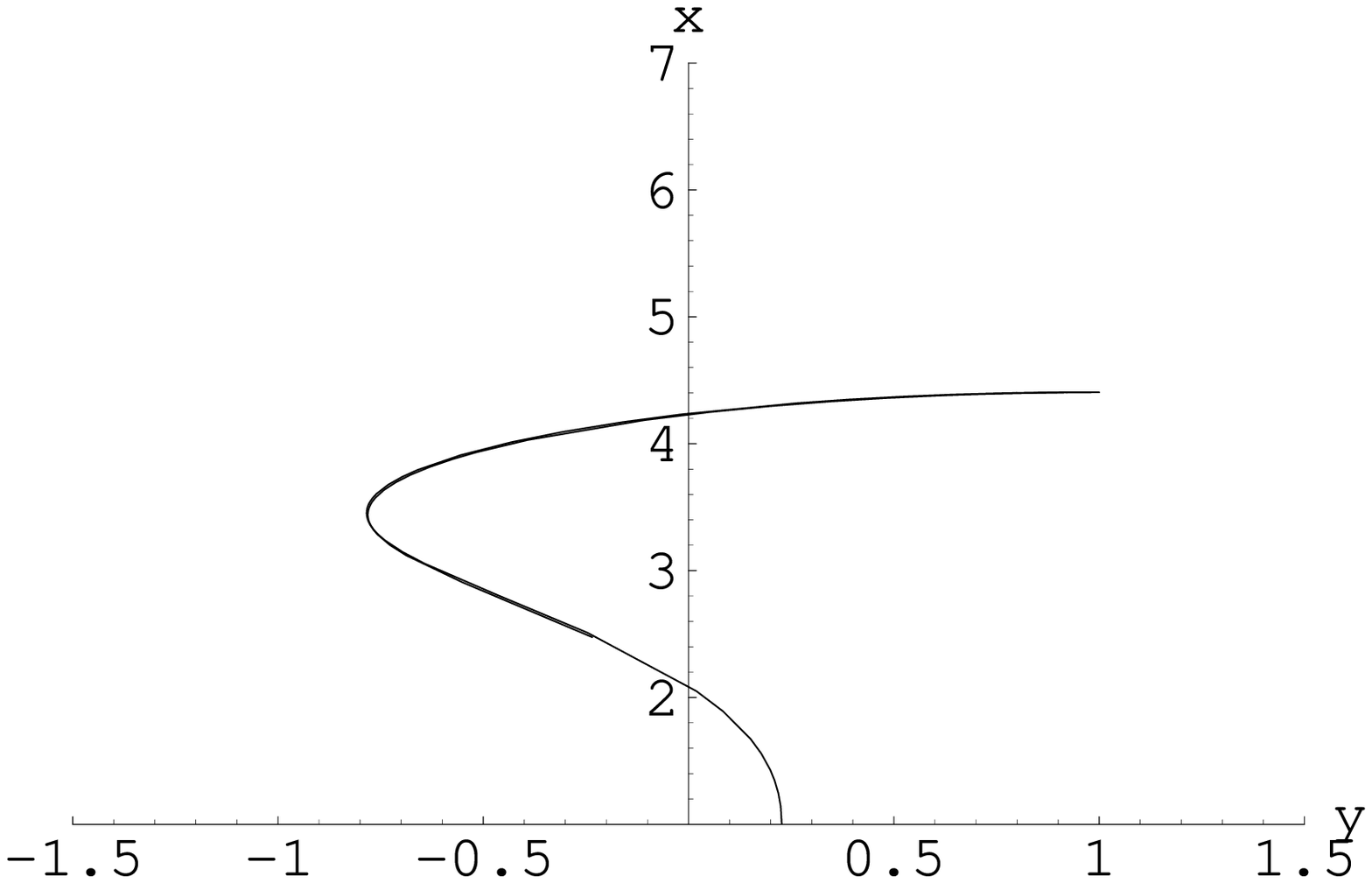}& 
\includegraphics[width=3in]{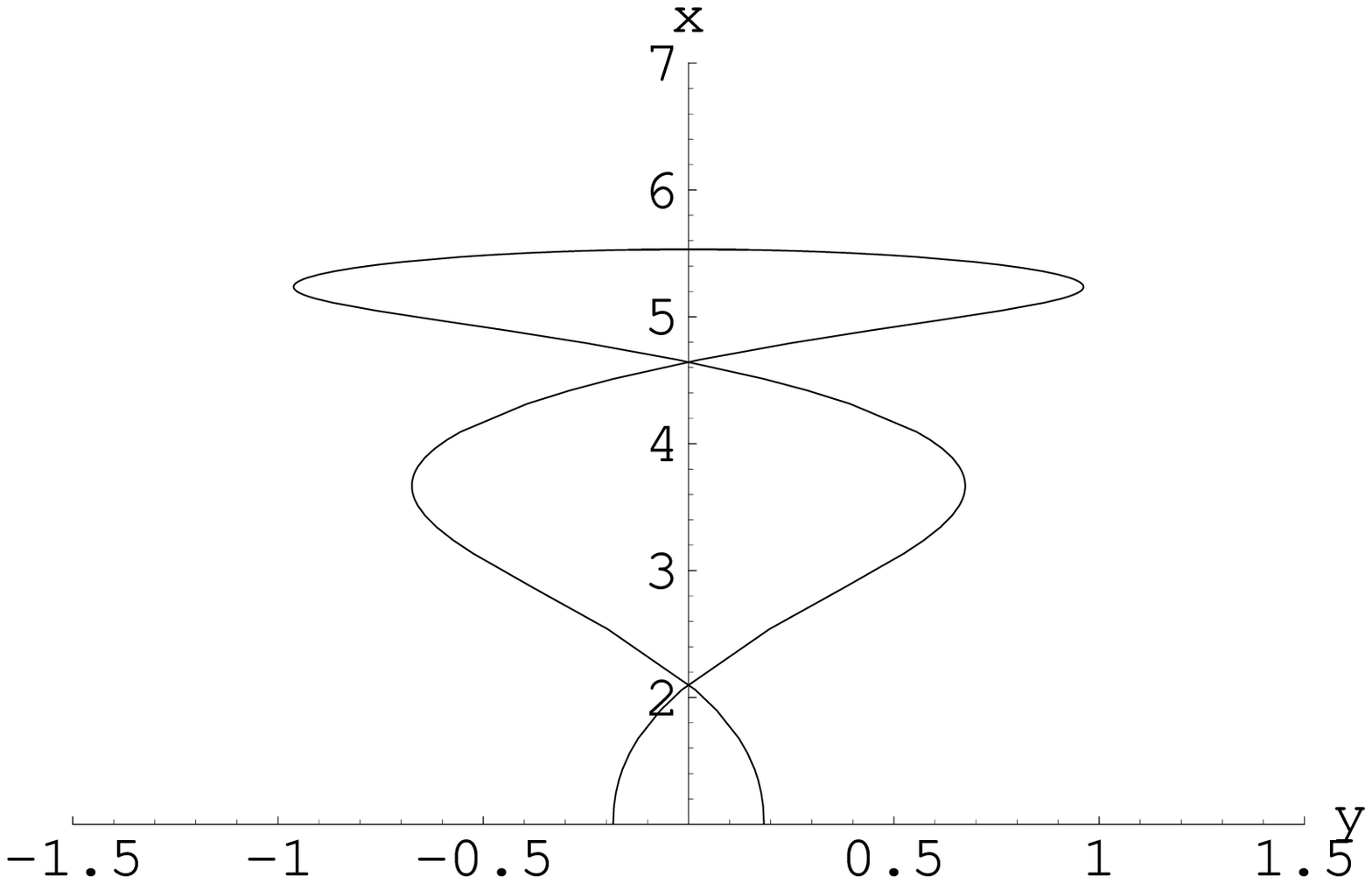} \\
(c) & (d) \\
\end{tabular}
\caption{Solutions to the equation of motion $(x(\sigma),y(\sigma))$ with initial conditions (a) $y(0) = 0.429$, (b) $y(0)=0.297$, (c) $y(0) = 0.227$, (d) $y(0) = 0.184$. \label{fige}}
\end{figure}

One obvious issue which we did not explore fully in section \ref{sec2}
is the meaning of other roots at $y(0)$ taking values 0.184, 0.227,
0.297, and 0.429 in figure \ref{figb}. It is straight forward to
illustrate the trajectory $(x(\sigma), y(\sigma))$ associated with
these solutions. They are illustrated in figure \ref{fige}. Each of
these solutions respects the boundary condition at both endpoints of
the string world sheet. As such, they also correspond to
positronium-like W/anti-W bound states of the theory which are
effectively stable in the large $N$, large $\lambda$ limit. For the
cases illustrated in figure \ref{fige}.a and \ref{fige}.c,
corresponding to $y(0)=0.429$ and $y(0)=0.227$, respectively, the
string hits $y(\sigma)=1$ where the piece of the string is moving at
the speed of light in $AdS_5$. At that point, one expects the string
to fold and return to the brane along the same path it took to get
there. For the cases illustrated in figure \ref{fige}.b and
\ref{fige}.d, the string self-intersects an integer number of times
before returning to the brane. While it is natural to expect a state
like this to decay in an interacting theory, this effect is suppressed
at leading order in the $1/N$ expansion.  Identical set of configurations were first found for the open strings ending on D7-branes in \cite{Kruczenski:2003be}

There is no reason not to expect this pattern of increasingly
self-intersecting states to exist as $y(0)$ approaches zero, although
they are harder to discover using a numerical algorithm simply because
the proper time of trajectory, parametrized by $\sigma$, becomes
large.  After all, they are precisely analogous to a different way of
shooting a pin-ball through the potential (\ref{pot}), such that it
crosses the $y=0$ axis as it rolls in the $x \gg 1$ region where there
is a shallow oscillating potential along the $y$ direction. The fact
that there is this rich spectrum of excited positronium-like bound
states is a prediction of ${\cal N}=4$ SYM at large $N$ and large
$\lambda$. Similar richness in the allowed string configurations was
noted also in \cite{Minahan:1998xq} where quark anti-quark potential
for ${\cal N}=4$ SYM in Coulomb branch was studied using the
techniques of AdS/CFT correspondence.

\section{Positronium in Magnetic Field}

Another simple extension of the analysis of section \ref{sec2} is
turning on a uniform magnetic field in the plane of rotation of the
positronium. This issue was originally investigated in the context of
open strings ending on D7-branes in \cite{Jensen:2008yp}. We will
repeat the analysis here because it is a useful warm-up exercise
before addressing the non-commutative case.

Let us begin by reviewing the semi-classical trajectory of the charges forming a bound state in a magnetic field.

\begin{figure}
\centerline{\includegraphics{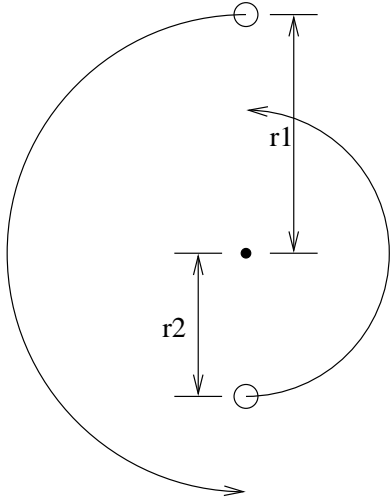}}
\caption{Circular orbit of charged particle anti-particle pair in a magnetic field.\label{figf}}
\end{figure}

A single moving charge in a magnetic field follows the standard circular orbit. If they are two opposite charges, they can orbit around a common axis with the same angular velocity $\omega$, and radius $r_1$ and $r_2$ as illustrated in figure \ref{figf}.

The equation of motion for the two charges are
\be m \omega^2 r_1 = {q^2\over (r_1+r_2)^2} + qB \omega r_1  \ee
\be m \omega^2 r_2 = {q^2 \over (r_1+r_2)^2} - q B \omega r_2  \ee
from which we infer
\be m \omega^2 R = {2 q^2 \over R^2}, \qquad m \omega^2 (r_1-r_2) = q  \omega R B, \qquad R = (r_1 + r_2)\ .  \ee
So we see that for fixed $\omega$ the effect of non-vanishing $B$ field is to shift the axis of rotation away from the center of mass without affecting the dipole length $R$.

Let us examine how the magnetic field affects the large $N$ large
$\lambda$ description in terms of semi-classical strings in $AdS_5
\times S_5$. The only change in the dual type IIB picture is that
there will be a constant NSNS 2-form potential along the plane of
rotation of the string. This will affect (\ref{action}) by adding a term
\be L_{B} = \int d \tau \, d \sigma \, \omega B r(\sigma) r'(\sigma)  \ . \ee
After the same change of variables as in the previous section, as well as scaling
\be B = {\sqrt{\lambda}U_0^2 b \over 2 \pi} \ee
this term becomes
\be L_{B} = {\lambda^{1/2} \over 2 \pi}  \int d \tau \, d \sigma \, b w  y(\sigma) y'(\sigma)  \ . \ee

Note that this term is a total derivative, and does not affect the
equation of motion. It does, however, modify the Neumann boundary
condition, when combined with (\ref{action2}), to read
\be y'(\sigma) = - b w y(\sigma) \label{bc2} \ee
at the boundaries of the open string world sheet $\sigma=0$ and
$\sigma=\sigma_{max}$. This boundary condition, combined with the
conservation of the Hamiltonian, also implies
\be x'(\sigma) = \pm\sqrt{{1 - w^2(1+b^2 x(\sigma)^4) y(\sigma)^2 \over x(\sigma)^4} } \label{xpbc} \ee
at the endpoints of the open string. We are, therefore, faced with a
similar problem of scanning over the set of initial condition
parametrized by $y(0)$ so that the boundary condition at
$\sigma=\sigma_{max}$ is satisfied. The only difference between this
and the analysis of the previous two sections is the change in the
boundary condition (\ref{bc2}).

\begin{figure}
\centerline{\includegraphics[width=3.5in]{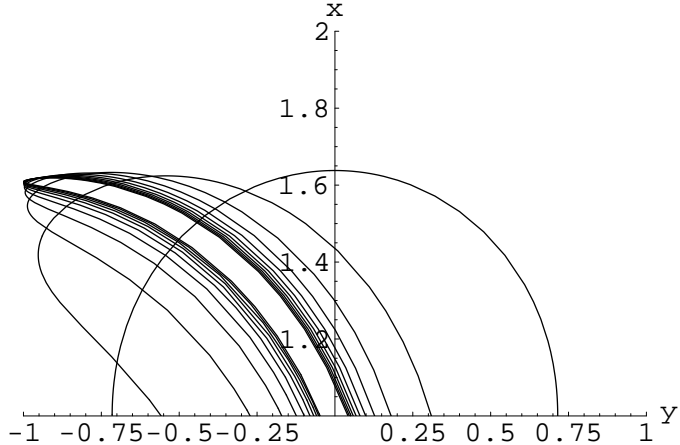}}
\caption{Configuration of spinning open string ending on a D3-brane at fixed $U_b$ in $AdS_5 \times S_5$ with non-vanishing $B_{23}$. Here, we consider $0 \le b  \le 10$ with $\omega=1.$
\label{figg}}
\end{figure}

For the sake of concreteness, we picked $x_b=1$, $w=1$, and scanned $b$ in the range $0 < b < 10$. The result of this analysis is illustrated in figure \ref{figg}.

Unlike what we found in the semi-classical analysis of the positronium
in a magnetic field, the length of the dipole does appear to shrink as
$b$ is increased. This is partly necessitated by the requirement that
(\ref{xpbc}) is real.  It is also interesting to note that the two
endpoints are located roughly symmetrically around the axis of
rotation. It is also interesting to note that as $B$ is increased, the
string appear to be approaching a configuration with a fold.

There is one more curious feature concerning the shape of the strings
illustrated in figure \ref{figg}.  As $b$ is varied keeping $\omega$
fixed, the maximum value of $x_{max}\approx 1.64$ of the trajectory
appears roughly constant.  Closer examination shows, $x_{max}$ does
vary mildly as $b$ is varied. We will encounter similar behavior for
the semi-classical open string configurations in the non-commutative theory.

\bibliography{wilson}
\bibliographystyle{utphys}

\end{document}